\documentclass[useAMS,usenatbib]{mn2e}

\usepackage{graphicx}
\usepackage{longtable}
\usepackage{pdflscape}
\usepackage{amsmath,amssymb}

\usepackage{color}

\begin{document}
\title{The Galaxy Cluster Concentration-Mass Scaling Relation} 
\date{}
\author[Groener, Goldberg, \& Sereno]
{A. M.~Groener,$^1$\thanks{Austen.M.Groener@Drexel.edu}
D. M.~Goldberg,$^1$
M. ~Sereno,$^2$ $^3$
\\
$^1$Department of Physics, Drexel University\\
Philadelphia, PA 19104\\
$^2$Dipartimento di Fisica e Astronomia, Alma Mater Studiorum Universit\`a di
Bologna, \\
viale Berti Pichat 6/2, 40127 Bologna, Italia\\
$^3$INAF, Osservatorio Astronomico di Bologna,\\
 via Ranzani 1, 40127 Bologna, Italia\\
}

\maketitle
\begin{abstract}
Scaling relations of clusters have made them
particularly important cosmological probes of structure formation. In this
work, we present a comprehensive study of the relation between
two profile observables,  concentration ($\mathrm{c_{vir}}$) and mass
($\mathrm{M_{vir}}$). We have collected the largest known sample of measurements from
the literature which make use of one or more of the following reconstruction
techniques: Weak gravitational lensing (WL), strong gravitational lensing (SL),
Weak+Strong Lensing (WL+SL), the Caustic Method (CM), Line-of-sight Velocity
Dispersion (LOSVD), and X-ray. We find that the concentration-mass (c-M)
relation is highly variable depending upon the reconstruction technique
used. We also find concentrations derived from dark matter only simulations
(at approximately $\mathrm{M_{vir} \sim 10^{14} M_{\odot}}$) to be
inconsistent with the WL and WL+SL relations at the $\mathrm{1\sigma}$
level, even after the projection of triaxial halos is taken into
account. However, to fully determine consistency between simulations and
observations, a volume-limited sample of clusters is required, as
selection effects become increasingly more important in answering
this. Interestingly, we also find evidence for a steeper WL+SL relation as
compared to WL alone, a result which could perhaps be caused by the varying
shape of cluster isodensities, though most likely reflects differences in
selection effects caused by these two techniques. Lastly, we compare
concentration and mass measurements of individual clusters made using
more than one technique, highlighting the magnitude of the potential bias
which could exist in such observational samples.

\end{abstract}

\section{Introduction}

Galaxy clusters have long been used as probes of cosmology. Cluster
observables, like X-ray luminosity, $\mathrm{L_{X}}$, optical
richness, and line-of-sight galaxy dispersion,
$\mathrm{\sigma_{v}}$, are closely tied to the formation and evolution of
large scale structures, and scale with redshift and the mass of the
host halo \citep{SE15.3}. Scaling relations of clusters also provide a way of testing
cosmology \citep{VI09.1,RO10.1,MA10.1,MA14.1}, though are imperfect
proxies for mass, due to the 2-Dimensional view they provide for us. Large
cosmological simulations provide a detailed 3-dimensional 
view of the hierarchical process of structure formation, one that is
unattainable by even the most accurate reconstruction techniques available.

The radial density profiles of clusters, well-modeled by the universal NFW
profile \citep{NA97.1}, appears to be a prevailing outcome of simulations
regardless of cosmology \citep{NA97.1,CR97.1,KR97.1,BU01.1}. 
\begin{equation}
\mathrm{\rho (r) = \frac{\delta_{c} \rho_{cr}}{\frac{r}{r_{s}}\left( 1 + \frac{r}{r_{s}}\right)^{2}}}
\end{equation}
\begin{equation}
\mathrm{\rho_{cr} = \frac{3 H(z)^{2}}{8\pi G}}
\end{equation}
However, the details of the relationship between the two model parameters, halo
mass, M, and concentration, $\mathrm{c}$, is sensitive to small changes in
initial parameters \citep{MA08.1,CO15.1}. The physical interpretation of a
halo's concentration (defined as the ratio of the virial radius,
$\mathrm{R_{vir}}$, to the radius at which $\mathrm{\rho\propto r^{-2}}$;
called the scale radius, $\mathrm{r_{s}}$), is that it is a measure of the
`compactness' of the halo, and determines the physical scale on which the
density profile rises steeply.


\setcounter{table}{0}
\begin{table*}
\begin{minipage}{210mm}
\caption{Population Overview}
\label{tab1}
\begin{tabular}{@{}lllcccllllll}
\hline
\hline
 Method &$\mathrm{N_{meas}}$&$\mathrm{N_{cl}}$&$\mathrm{min(M_{vir}/10^{14}M_{\odot})}$
              &$\mathrm{\langle M_{vir}/10^{14}M_{\odot} \rangle}$ 
 &$\mathrm{max(M_{vir}/10^{14}M_{\odot})}$ &$\mathrm{min(c_{vir})}$&$\mathrm{\langle c_{vir} \rangle}$ 
 &$\mathrm{max(c_{vir})}$& $\mathrm{min(z)}$ & $\mathrm{\langle z
                                                     \rangle}$
                            &$\mathrm{max(z)}$\\ 
\hline
CM & 82 & 79 & $<$1.0 & 3.9 & 18.6 & $<$2.0& 8.9 &36.7 & 0.003& 0.06&0.44\\
LOSVD & 70 & 59 & 1.3& 5.8 & 17.1& $<$2.0& 8.8& 39.0& 0.01& 0.06&0.44\\
X-ray & 290& 195 & $<$1.0 & 26.1 & $>$40.0& $<$2.0& 7.2& 26.2& 0.003 & 0.22&1.41\\
WL & 169 & 111 & $<$1.0 & 12.4 & $>$40.0& $<$2.0& 8.1& 64.5& 0.02& 0.48 &1.45 \\
WL+SL & 113 & 58 & $<$1.0 & 8.7 & 31.8 & 2.3& 10.2& 30.6& 0.18& 0.53 &1.39\\
SL & 19 & 11 & 3.2& 24.3 & $>$40.0 & 3.8& 11.2& 27.5& 0.18& 0.47 &0.78\\
\hline
\hline
\end{tabular}
\end{minipage}
\end{table*}

The first indication of the connection between halo concentration and mass
(hereafter, the c-M relation) was discovered through simulations of structure
formation by \citet{NA97.1}, and later confirmed by \citet{BU01.1}, who found a
strong correlation between an increasing scale density, $\mathrm{\rho_{s}=\delta_{c}
  \rho_{cr}}$, for decreasing mass, $\mathrm{M_{vir}}$. The explanation for
this anti-correlation between concentration and mass is that low-mass halos
tend to collapse and form relaxed structures earlier than their larger
counterparts, which are still accreting massive structures until much later. A
consequence of early collapse is that halos will have collapsed during a period
of higher density, leading to a larger central density (and hence larger
concentration) as compared to halos which formed later.

Many studies (most recently, e.g., \citealt{CO15.2}) focus on the physical motivation
for the existence of this relationship, and suggest that the mass
accretion history (MAH) of halos is the key to understanding the connection
between cluster observables and the environment in which they formed
\citep{BU01.1,WE02.1,ZH03.1}. These studies have found that while the
mass accretion rate onto the halo is slow, the concentration tends to scale with the virial
radius, $\mathrm{c\propto r_{vir}}$ (caused by a constant scale radius), while
the concentration remains relatively constant for epochs of high mass
accretion. The MAH itself depends upon the physical properties of the initial
density peak \citep{DA08.1}, which is a function of cosmology, redshift, and
mass \citep{DI15.1}.

Longstanding tension has existed between cluster concentrations derived from
simulations and observational measurements. Concentrations have been found to
differ the most for gravitational lensing techniques
\citep{CO07.1,BR08.1,OG09.1,UM11.2}. This over-concentration in favor of
observational measurements can be partially explained by the orientation of
triaxial structure along our line-of-sight \citep{OG05.1,SE11.1}, which has the
effect of enhancing the lensing properties \citep{HE07.1}. Neglecting halo
triaxiality \citep{CO09.1} and substructure \citep{ME10.2,GI12.1} also each
have significant effects on halo parameters. For its effect on WL and X-ray
mass estimates, see \citet{SE14.2}.

Discrepancies in how measurements of the intrinsic concentration
are made using simulations also exist, along with studies who disagree
on the inner slope of the density profile
\citep{MO99.2,GH00.1,NA04.1}. However, the most puzzling and potentially
interesting disparity between simulations is the existence of the upturn
feature in the c-M relation (see for example, Fig. 12 of \citealt{PR12.1}) at high redshift
\citep{PR12.1,DU14.1,KL14.1,DI15.1}, which some argue is an artifact caused by
the selection of halos which are dynamically unrelaxed \citep{LU12.1}. This
novel feature only shows up when the concentration is expressed as a
profile-independent halo property (in terms of the ratio of the maximum 
circular velocity and the virial velocity, $\mathrm{V_{max}/V_{vir}}$). In
terms of the classical definition of concentration, this feature disappears
(see \citealt{ME13.1}). 

The connection between the observed concentration, $\mathrm{c_{2D}}$,
and the intrinsic concentration, $\mathrm{c_{3D}}$, is further complicated,
since it has been shown that relaxed cluster isodensities are not constant
on all scales \citep{FR88.1,CO96.1,DU91.1,WA92.1,JI02.1,HA07.1,GR14.1}. Indeed,
in a previous study by \citet{GR14.1}, it has been shown that a halo's concentration
is an ill-defined 2-dimensional quantity, without first specifying the scale on which the
measurement was made. Using the MultiDark MDR1 Cosmological
Simulation, \citet{GR14.1} found a systematic shift of about $\mathrm{\sim 18 \%}$ in the
mean value of the projected concentration, $\mathrm{c_{2D}}$, between weak and
strong lensing scales, for low-mass cluster halos ($\mathrm{2.5-2.6 \times
  10^{13} h^{-1} M_{\odot}}$) observed with their major axes aligned with the
line-of-sight direction. Though this difference is notably smaller than the intrinsic
scatter of the concentration parameter ($\mathrm{c_{3D}}$) for a given halo
mass, the origin of this systematic effect is solely due to the changing shape
of cluster isodensities as a function of radius.

For many objects, not only do observed concentrations seem to differ
substantially from those obtained in cosmological simulations, but
concentrations can also vary depending on which method is used. Since different
reconstruction methods probe varying scales within the halo, it is 
not unreasonable to suspect that there exist systematic differences in the
observed c-M relation caused by shape.

\bigskip

\noindent In this paper, we focus on three main objectives.

\begin{enumerate}
\item We present the current state of the observational concentration-mass
  relation for galaxy clusters by aggregating all known measurements from the
  literature. The raw data are reported in Table A-1, and have been made
  publicly available (see Appendix A). We also provide an additional table
  (available only online), where data have been normalized over differences in
  assumed cosmology, overdensity convention, and uncertainty type found in the
  original studies.

\item We model the observed concentration-mass relation for each method,
  and compare these to one another, highlighting potential differences which
  exist, caused by the projection of structure along the
  line-of-sight, the varying shape of cluster isodensities, and the selection
  of clusters from the cosmic population.

\item Using the largest cluster sample to date, we determine
  if the observed c-M relation is consistent with theory, when
  taking halo triaxiality and elongation of structure along the line-of-sight
  into account.
\end{enumerate}

In section 2, we summarize many of the most common mass reconstruction
techniques which are used throughout the cluster community, and include a
discussion regarding physical scales probed within the cluster using these
methods. In section 3, we discuss the procedure for collecting our sample from
the literature, and normalizing over convention, cosmology, and
uncertainties. In section 4, we present results for the observed c-M relation
for each method, and in section 5, we discuss the projection of triaxial halos
from simulations to the observed lensing relations. Lastly, in section 6, we
conclude and discuss our findings.

Throughout this paper we adopt a flat $\Lambda$CDM cosmology, $\Omega_{m}=0.3$,
$\Omega_{\Lambda}=0.7$, and $\mathrm{H}_{0} = 70$ $\mathrm{km}$
${\mathrm{s}}^{-1} {\mathrm{Mpc}}^{-1}$. Generally speaking, we reserve the
following colors within plots to represent the various methods: 
\begin{itemize}
\item Caustic Method (CM): blue
\item Line-of-sight velocity dispersion (LOSVD): orange
\item X-ray: green
\item Weak Lensing (WL): purple
\item Strong Lensing (SL): red
\item Weak + Strong Lensing (WL+SL): black
\end{itemize}
Unless otherwise stated, throughout the study, uncertainties are reported as
1-$\mathrm{\sigma}$ (68.3\%) Gaussian uncertainties.


\section{Cluster Mass Reconstruction Techniques}
In this section, we present a brief overview of common mass reconstruction
techniques and modeling of the cluster density profile.
\subsection{Weak Lensing (WL)}
Weak gravitational lensing is the process by which images of background galaxies
are distorted by massive foreground objects. Though these distortions cannot be
detected for any given source, it is possible to obtain a signal by
locally averaging the shapes (ellipticities) of galaxies. This shear measurement within a given
bin can be used as a direct proxy for the lens density profile at intermediate to large radii.


For a symmetric distribution, the azimuthally averaged tangential shear,
$\mathrm{\langle \gamma_{t} \rangle}$, as a function of radius from the cluster
center can then be calculated, and relates to the convergence, $\kappa$, in the
following way: 
\begin{equation}
\mathrm{\langle \gamma_{t} \rangle (r) = \frac{\bar{\Sigma}(<r) -
    \bar{\Sigma}(r)}{\Sigma_{cr}} = \bar{\kappa}(<r) - \bar{\kappa}(r) }
\end{equation}
where the critical surface mass density is defined in terms of
cosmology-dependent angular diameter distances $\mathrm{D_{s}}$ (source),
$\mathrm{D_{ds}}$ (lens to source), and $\mathrm{D_{d}}$ (lens): 
\begin{equation}
\mathrm{\Sigma_{cr} = \frac{c^{2} D_{s}}{4\pi G D_{ds} D_{d}}}
\end{equation}
Expressions, specifically for the NFW profile, for the convergence
\citep{BA96.1} and the tangential shear \citep{WR99.1} have been
derived, and can be used for model fitting.

Weak lensing comes with its own intrinsic biases in that more massive clusters
produce larger distortions of background galaxies. As a result, in a survey of
clusters, the expectation is that nearly all of the most massive clusters would
be {\em selected} from the population. However, in the low mass region,
clusters which are highly triaxial and elongated along the line-of-sight (i.e. -
larger 2D concentrations) are more likely to pass the observational
signal-to-noise threshold than ones which are not. The net effect here is an
artificial steepening of the c-M relation due to selection. Furthermore,
lensing geometry plays an additional role in how clusters are
selected. Clusters which are too distant lack the requisite number 
density of background galaxies to obtain high signal-to-noise
\citep{BA01.1}. Table 1 presents the range in redshift for weak lensing
clusters, where most measurements are found to lie in the redshift range of
$\mathrm{z=0.2-0.6}$, with $\mathrm{M_{vir} \gtrsim 1\times 10^{14}
  M_{\odot}}$.

\begin{figure}
\begin{center}
\includegraphics[scale=0.4]{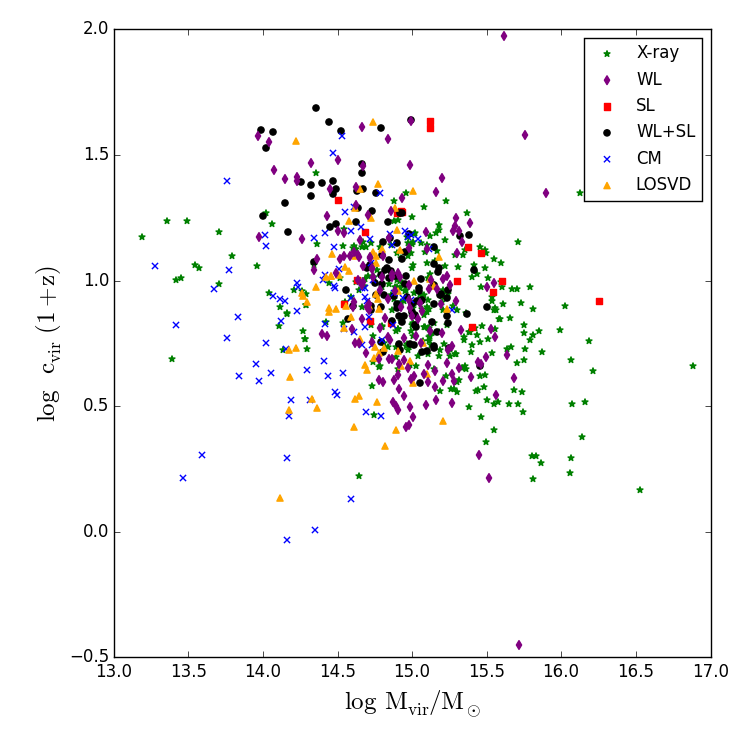}
\end{center}
\caption{The full normalized observational cluster sample, colored by
  method. Uncertainties have been omitted here for clarity.}
\end{figure}

\subsection{Strong Lensing (SL)}
A natural extreme of the phenomenon of gravitational lensing can occur
if a background galaxy is serendipitously aligned with the core of
a cluster. In such cases, the projected surface mass density is so high that
multiple images of the object are produced, commonly distorting them so much
that they appear arc-like.

A density profile can be obtained by fitting a model to the observed image
positions, orientations, and fluxes, though this technique
constrains the cluster profile on small scales (approximately the Einstein radius,
$\mathrm{ \theta_{E}}$\footnote[1]{The Einstein radius for a point mass is
  $\mathrm{\theta_{E} = \left( \frac{4GM}{c^{2}} \frac{D_{LS}}{D_{L}D_{S}}
    \right)^{1/2}}$. Though there is no corresponding functional form for an
  NFW profile,  typical values for clusters lie in the range: 10''-45''
  \citep{KN03.1,BR05.2}.}, which is typically $\mathrm{\sim 5\%}$ of the virial
radius, $\mathrm{r_{vir}}$, or $\mathrm{\sim 50\%}$ of the scale radius,
$\mathrm{r_{s}}$ \citep{OG09.2}).

Due to the irregular occurrence of multiple images and arcs, cluster
measurements made with strong lensing are particularly prone to selection
effects, and likely represent a biased sampling of the cosmic
population. In fact, the efficiency of lensing is increased with increasing
mass and concentration, and a preferential line-of-sight alignment of the
triaxial halo \citep{OG09.2}. Concentrations derived from this method have been
contentiously high as compared to X-ray studies \citep{CO07.1}.

\subsection{Weak+Strong Lensing (WL+SL)}
Combining weak and strong gravitational methods constrains the density
profile over a wide range of scales, and also has the ability to break the
mass-sheet degeneracy \citep{SC95.1}. Recent efforts to combine these methods
have become more prevalent in the literature (\citet{ME14.1} - 
CLASH; \citet{OG12.1} - SGAS), and work to reconstruct the lensing potential by
minimizing a combined least-squares approach. 
\begin{equation}
\mathrm{\phantom{.} \chi^{2}(\psi) = \chi^{2}_{w}(\psi) + \chi^{2}_{s}(\psi)}.
\end{equation}

\subsection{X-ray}
Massive clusters are significant sources of X-ray radiation, due to the hot
diffuse plasma ($\mathrm{k_{B}T_{e} \sim 10 \, keV}$) emitting via thermal
bremsstrahlung, and can be used to determine the total distribution of
mass. Under assumptions of spherical symmetry and hydrostatic equilibrium with
the underlying potential \citep{EV96.1}, temperature and gas density
information, $\mathrm{\rho_{g}}$, are used to determine the total mass of the
cluster, typically at intermediate scales ($\mathrm{\sim r_{500} }$,
corresponding to the radius at which the average density inside is 500 times
$\mathrm{\rho_{cr}}$).
\begin{equation}
\mathrm{M(r) = \frac{k T(r)}{G\mu m_{p}} r \left( \frac{d \log \rho_{g}(r)}{d \log
  r} + \frac{d \log T(r)}{d \log r} \right) }
\end{equation}

These assumptions are often violated due to non-thermal pressure sources,
temperature inhomogeneity, and to the presence of substructures further out
\citep{RA12.1}, and bias low mass estimates by 25-35\%.

\subsection{Line-of-sight Velocity Dispersion (LOSVD)}
The distribution of mass within clusters can also be obtained by using the
kinematics of cluster galaxies, specifically, by using the moments of
the velocity distribution. Reconstruction methods, developed by
\citet{LO02.1} and \citet{LO03.1}, use the second (dispersion) and fourth
(kurtosis) moments of the velocity distribution, which
relies on the  underlying gravitational potential. Assuming the distribution of
mass follows an NFW profile, free parameters, which include $\mathrm{M_{vir}}$
and $\mathrm{c_{vir}}$, can be fit to the observed data.

The business of identifying clusters as mass over-densities, determining
cluster membership, removal of interlopers, and reconstruction details vary from
technique to technique. For a more complete review of the reconstruction
methods and their impact on cluster observables, see \citet{OL14.1}.

\subsection{The Caustic Method (CM)}
With the exception of weak lensing, the caustic method is the only other
standalone method which has been successful in probing the density profile at
large distances from the cluster center ($\mathrm{\gtrsim r_{vir}}$). Cluster
galaxies, when plotted in line-of-sight velocity versus projected
cluster-centric distance phase-space, create a characteristic ``trumpet
shape'', the boundaries of which form what is referred to as caustics
\citep{KA87.1,RE89.1}. The existence of these caustics mark an important
boundary which envelops a volume of space in which galaxies are gravitationally
bound to the cluster. Outside of this turnaround radius, galaxies are
ultimately carried away in the Hubble flow.

The width of the caustic (velocity) at any given projected radius,
$\mathcal{A}\mathrm{(R)}$, can then be related to the escape velocity due to
the gravitational potential of the cluster, under the assumption of spherical
symmetry \citep{DI97.1}. Through simulations of structure formation,
\citet{DI99.1} has shown that the caustic amplitude can be related to the mass
interior to radius $\mathrm{r}$ by: 
\begin{equation}
\mathrm{GM(<r) = \frac{1}{2} \int_{0}^{r} \mathcal{A}^{2}(R) \, dR }.
\end{equation}

The success of the caustic method is independent of any assumptions regarding
dynamical equilibrium of the cluster, and has been used to reconstruct profiles
over a larger range of scales: from the inner regions to a few times the virial
radius (CAIRNS: \citealt{RI03.1}; CIRS: \citealt{RI06.1}; HeCS:
\citealt{RI13.1}). However, this technique requires the measurements of at
least 30-50 cluster members, and thus limits this method to clusters at relatively low
redshifts compared to lensing and X-ray techniques. More recently,
\citet{RI13.1} make use of this technique using $\mathrm{\sim 200}$ cluster
members.

\subsection{Hybrid Techniques}
The aforementioned methods represent the most commonly applied techniques for
constructing a density profile, however, they do not represent them all. Novel
combinations of methods have also been used, but could not be included in a
study of this kind. For instance, \citet{LE08.1} combine joint lensing and
X-ray methods to make a determination of Abell 1689. \citet{TH10.1} and
\citet{VE11.1} use a combination of lensing and dynamics. Additionally, In an
attempt to only compare methods used in \citet{CO07.1}, we consciously
leave out measurements made with the Sunyaev-Zel'dovich (SZ) effect, or which use
combinations of techniques one of which uses SZ.

Previous studies have even employed these multi-technique reconstructions to
clusters in an attempt to break the line-of-sight mass degeneracy (for a review
of these techniques, see section 2 of \citet{LI13.1}; see also \citet{AM07.1},
\citet{SE12.1}). However, it is unclear if techniques such as this can adapt to 
arbitrarily complicated profiles, where shape is scale-dependent, or where
isodensities are not co-axial with one another (isodensity twisting).

\begin{figure*}
\begin{center}
\includegraphics[scale=0.6]{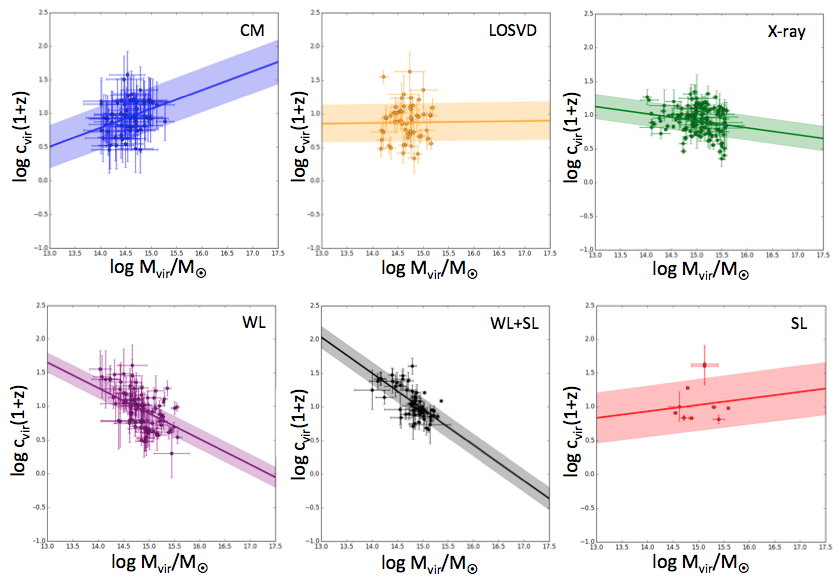}
\end{center}
\caption{Upper left to lower right: Individual fits to CM, LOSVD, X-ray,
    WL, WL+SL, and SL. The shaded regions represent the 1-$\mathrm{\sigma}$ uncertainty
    in the best-fit parameters, and includes the intrinsic scatter, $\mathrm{\sigma_{int}}$. These
    relations are extrapolated over the full range of cluster masses for
    illustration purposes only. }
\end{figure*}

\section{The Sample}

The sample of clusters collected from the literature consists of a
total of 781 cluster measurements, reported by 81 studies (Table A-2), representing the
largest known collection of cluster concentration measurements to date. Of
these, there are 361 unique clusters, giving us a sizable sampling of
the cluster population as a whole, in addition to multiple measurements of
individual clusters (often coming from more than one category of
reconstruction technique).

This study builds off of work done by \citet{CO07.1}, which aggregated 182
cluster measurements of 100 unique cluster objects. In accordance with that
study, we also report measurements of concentration (and mass) in the most
popular conventions, $c_{200}$, and $c_{vir}$. 

Table 1 presents population averages of masses and concentration, as well as
their range in redshift for the six reconstruction techniques we reference
throughout this study. This information highlights the importance of the
selection function of clusters, though we make no attempt in this paper to
distinguish between whether a lack of measurements of certain values for a given
method is due to its inability to make these determinations, or whether it is
simply a preferential selection effect.

\subsection{Normalization Procedure}

Due to the nature of this study, cluster measurements must be properly
normalized to ensure that they are compared to one another on equal footing. In
this section, we discuss the steps taken to eliminate biases due to overdensity
convention, assumed cosmology, and due to differences in the definitions of
measurement uncertainty, respectively. 

\subsubsection{Convention}
Under the assumption that the radial density profile follows an NFW profile,
\citet{HK03.1} derive a procedure for the conversion of both concentration and
mass between any two arbitrary characteristic radii. We apply these 
formulae as a first round of our normalization procedure.
\subsubsection{Cosmology}
Measurements taken from the literature do not always use the same
fiducial cosmology, and thus are not immediately comparable. Because of this,
we develop a procedure for converting measurements between any two arbitrary
cosmologies. Appendix B outlines this procedure for general lensing methods.

For extreme cosmologies, the correction to the concentration parameter,
$\mathrm{c_{vir}}$, and mass, $\mathrm{M_{vir}}$, are approximately 5$\%$ and
10$\%$, respectively. This correction is significantly smaller than other known
effects. Moreover, the vast majority of all measurements we have collected
assume flat cosmologies which lie in the range $\mathrm{\Omega_{\Lambda} = 1 -
  \Omega_{m}  = 0.73 - 0.68}$. The corrections to the concentration and mass in
this range are $\mathrm{\sim 1\%}$.

\subsubsection{Uncertainties}
Another complication which must be accounted for is the usage of
multiple definitions of measurement uncertainty on resulting mass
and concentration estimates reported throughout the literature. Particularly,
many fitting procedures (namely methods which involve brute force exploration
of likelihood space) produce maximum-likelihood estimates of
parameters of interest and corresponding confidence intervals. However, most
studies do not report the marginal distributions from their fitting procedures,
and consequently, limits the utility of their measurements for those looking to
compare or adopt their values.

Furthermore, the mathematical theorems which dictate the propagation
of error of measurements rely on expected values and variances, rather
than maximum-likelihood estimates and probability
intervals. \citet{DA04.1} argues that the expected value and standard
deviation should {\em always} be reported, and in the event of an
asymmetric distribution, one should also report shape parameters or
best-fit model parameters as well. Most importantly, any published
result containing asymmetric uncertainties causes the value of the
physical quantity of interest to be biased.

We follow the procedure outlined in \citet{DA04.1} for symmetrizing
measurements with asymmetric uncertainties (to first order),
$\mathrm{{\theta_{m}}^{\Delta_{+}}_{\Delta_{-}}}$, and apply this to both
cluster mass and concentration measurements.

\begin{subequations}
\begin{align}
\mathrm{\sigma_{\theta} \approx \frac{\Delta_{+} + \Delta_{-}}{2}}\\
\mathrm{E[\theta] \approx \theta_{m} + \mathcal{O}(\Delta_{+} - \Delta_{-})}
\end{align}
\end{subequations}

\begin{table*}
\begin{minipage}{210mm}
\caption{Best-Fit Concentration-Mass Relation Parameters}
\label{tab2}
\begin{tabular}{@{}llllllllllllllll}
\hline
\hline
 Method & & & & & & & & & & \multicolumn{3}{l}{Bootstrap $\rightarrow$}& & & \\ 
\hline
 &$\mathrm{N_{cl}}$&$\mathrm{m}$\footnote[1]{The slope, $\mathrm{m}$, of the
 linear model is exactly equivalent to the power-law index
 $\mathrm{\alpha}$.}&$\mathrm{\sigma_{m}}$\footnote[2]{$\mathrm{\sigma_{m} =
                      \sigma_{\alpha}}$} &$\mathrm{b}$ 
                &$\mathrm{\sigma_{b}}$ &$\mathrm{A}$\footnote[3]{The
                                         normalization parameter, $\mathrm{A}$,
                                         depends upon both $\mathrm{m}$ and $\mathrm{b}$:
                                         $\mathrm{A=10^{b+m\log M_{*}}}$}
                    &$\mathrm{\sigma_{A}}$\footnote[4]{Uncertainty was
                      propagated through the expression in [3].}
                    &$\mathrm{\sigma_{int}}$\footnote[5]{Equivalent to the
                      scatter in $\mathrm{\log c_{vir}}$ reported in previous studies.}& $\mathrm{\chi^{2}_{red}}$& $\mathrm{m}$ &$\mathrm{\sigma_{m}}$ &$\mathrm{b}$
                            &$\mathrm{\sigma_{b}}$ &$\mathrm{A}$
                                                          &$\mathrm{\sigma_{A}}$\\
\hline
CM & 63 & 0.280 & 0.003 & -3.138 & 0.038 & 3.778 & 0.677 & 0.242 & 0.327 & 0.28& 0.19& -3.16&2.73 & 3.59&43.43\\
LOSVD & 58 & 0.010 & 0.002 & 0.728& 0.025 & 7.256 & 0.861 & 0.228 & 1.000 & 0.13& 0.17& -1.00& 2.55& 5.31&58.74\\
X-ray & 149 & -0.105 & 0.001 & 2.494 & 0.010 & 12.612 & 0.676 & 0.160 & 1.224 &-0.17 & 0.03& 3.38& 0.44& 13.32&25.69\\
WL & 93 & -0.379 & 0.001 & 6.576 & 0.014 & 35.246 & 2.213& 0.118 & 1.302 & -0.43& 0.11& 7.35& 1.62& 44.10&312.68\\
WL+SL & 57 & -0.534 & 0.001 & 8.977& 0.016 & 77.882 & 5.249 & 0.130 & 1.070 & -0.54& 0.10& 9.10& 1.46& 86.06&552.28\\
SL & 10 & 0.097 & 0.004 & -0.422 & 0.062 & 7.236 & 1.951 & 0.254 & 1.003 & 0.11& 0.23& -0.60& 3.49& 7.24&109.02\\
\hline
All (This Work) & 293 & -0.152 & 0.001 & 3.195 & 0.007 & 15.071 & 0.703 & 0.146 & 1.354 & -0.16& 0.03& 3.26& 0.44& 13.71&26.45\\
All (CO07.1)& 62 & -0.14 & 0.12 & -- & -- & 14.8 & 6.1 & 0.15 & -- & -- & -- & -- & -- & -- & --\\
\hline
\hline
\end{tabular}
\end{minipage}
\end{table*}

Additionally, many studies report measurements without uncertainties
altogether. For these clusters, we apply uncertainty based upon the
estimate of the average fractional uncertainty of all other
measurements of its type. The most notable method having this issue is
the caustic method, where virtually no measurements are accompanied by
uncertainties. In this case, we apply the same fractional uncertainty
to all measurements equally, and is derived from the average fractional error
of LOSVD concentration and mass measurements.

Lastly, a large fraction of clusters represented in our database have multiple
concentration and mass measurements, leading subsequent fits to be more
sensitive to these particular objects. In order to prevent fits from being
dominated by the most popular clusters (e.g. - Abell 1689, of which there are
26 measurements in total), we combine similar measurements using an
uncertainty-weighted average value.

\section{The Observed Concentration-Mass Relation}

In Figure 1, we show the full cluster dataset after applying the normalization
procedures discussed in the previous section. Following this,
we present here the results of our fitting procedure to these data. The typical
prescription for modeling the c-M relation, is to use a double power-law model
of the following form
\begin{equation}
\mathrm{c(M) = \frac{A}{\left(1+z\right)^{\beta}} \left(\frac{M}{M_{*}}\right)^{\alpha}}
\end{equation}
where the power-law indices, $\mathrm{\alpha}$ and $\mathrm{\beta}$, control
the dependence of the concentration with respect to mass and redshift. The
model parameter, $\mathrm{A}$, controls the normalization of the relation, once a
suitable $\mathrm{M_{*}}$ has been chosen ($\mathrm{M_{*}=1.3\times10^{13}
  h^{-1} M_{\odot} = 1.857\times 10^{13} M_{\odot}}$). 

We follow convention in using the above model, but in a slightly different
form, with the power-law index, $\mathrm{\beta}$, fixed to unity. The
particular choice of $\mathrm{\beta=1}$, and pivot mass
$\mathrm{M_{*}=1.3\times10^{13} h^{-1} M_{\odot}}$, is for ease of comparison
with previous large studies of the c-M relation (\citealt{CO07.1}). We adapt
this model to a linear model in the following way
\begin{equation}
\mathrm{\mathcal{Y} = m\mathcal{X}+b \pm \sigma_{int}}
\end{equation}
where variables and model parameters relate to the initial model in the
following way:
\begin{subequations}
\begin{align}
\mathrm{\mathcal{Y}}&\mathrm{\equiv \log c (1+z)}\\
\mathrm{\mathcal{X}}&\mathrm{\equiv \log M}\\
\mathrm{m}&\mathrm{= \alpha}\\
\mathrm{b}&\mathrm{= \log A - \alpha \log M_{*}}
\end{align}
\end{subequations}

We introduce the intrinsic scatter, $\mathrm{\sigma_{int}}$, as a fixed parameter,
which we estimate from the data (independently from the fit
itself), and is assumed to be constant over the full mass range:
\begin{equation}
\mathrm{\sigma_{int}^{2} = \sigma_{res}^{2} - \langle \sigma_{\mathcal{Y}}^{2} \rangle}
\end{equation}
where $\mathrm{\sigma_{res}}$ is the scatter in the residual between the data and the
best-fit model, and $\mathrm{\langle \sigma_{\mathcal{Y}}^{2} \rangle}$ is the
average squared-uncertainty in the dependent variable. The idea here is that the
scatter in the residual must be accounted for by a combination of scatter due
to the intrinsic relation itself as well as the uncertainties in the
measurements of the observables. We also note that although the value of the
redshift for any given cluster has an effect on the uncertainty of the variable
$\mathrm{\mathcal{Y}}$, the uncertainty in the measured redshifts themselves do
not contribute much to the overall uncertainty of the best-fit model parameters.
 
After measurements have been normalized, we eliminate extreme values of mass
and concentration. Simulations tell us that the most massive clusters which
exist at present are approximately a few times $\mathrm{10^{15} \,
  M_{\odot}}$. Accordingly, we remove masses which are larger than
$\mathrm{4\times 10^{15}\,M_{\odot}}$. We also remove masses lower than
$\mathrm{1\times 10^{14}\, M_{\odot}}$, since best fit parameters are
particularly sensitive to this mass bin (representing data for low-mass galaxy
clusters and galaxy groups). Lastly, concentrations which are lower than 2,
indicate rather poor NFW fits to the density profile, and will bias our
inferred parameters.

In Table 2, we present our best-fit linear model parameters, and their mapping
back to the original power-law model. In Figure 2, individual fits
to each subsample are shown alongside normalized data points. Lensing (WL
and WL+SL) and X-ray relations show a clear trend consistent with concentration
decreasing with increasing mass. We also include a bootstrap analysis of
these fits, to reveal the sensitivity of the fits to the data. 

Though seemingly well-constrained, the bootstrap analysis reveals that our
strong lensing c-M relation is highly sensitive to the dataset (due to the very
small sample size), and so the best-fit model parameters are likely
untrustworthy.

General agreement between concentration and mass measurements of all methods
can be seen in the range $\mathrm{10^{14.5}-10^{15} M_{\odot}}$, which we also point
out, is the region we find most consistent with simulation results.

\begin{figure*}
\begin{center}$
\begin{array}{c}
\includegraphics[width=0.75\textwidth]{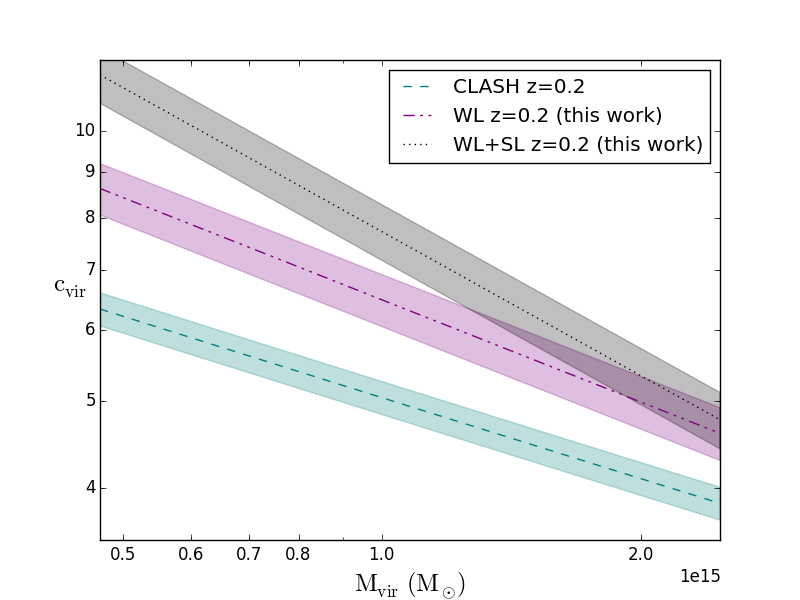} \\
  \includegraphics[width=0.75\textwidth]{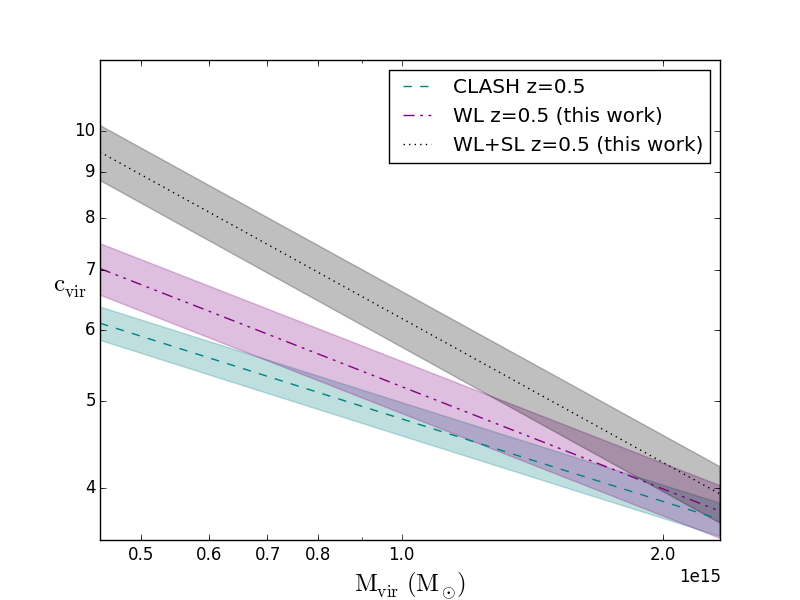}
\end{array}$
\end{center}
\caption{A direct comparison of the concentration-Mass relations for
  lensing based methods (WL and WL+SL) with results from CLASH
  \citep{ME14.1}. The {\em top} panel shows these relations at a redshift of
  $\mathrm{z=0.2}$, whereas the {\em bottom} panel is at a higher
  redshift $\mathrm{z=0.5}$ (approximately the average redshift of WL
  and WL+SL measurements in our sample). Conversion from $\mathrm{c_{200}}$
  to $\mathrm{c_{vir}}$ was necessary for comparison purposes.}
\end{figure*}

We also compare our results to the c-M relation studied by CLASH, which use
a combined weak and strong lensing technique for 19 X-ray selected galaxy
clusters. The relation they fit,
\begin{equation}
\mathrm{c_{200} = A \left( \frac{1.37}{1+z} \right)^{B} \left(
    \frac{M_{200}}{8\times 10^{14} h^{-1} M_{\odot}} \right)^{C}}
\end{equation}
with best-fit values of $\mathrm{A=3.66\pm0.16}$, $\mathrm{B = -0.14\pm0.52}$,
and $\mathrm{-0.32\pm0.18}$, agrees well with projected simulations, after
accounting for the X-ray selection function. Figure 3 shows the comparison of
the CLASH c-M relation to the the lensing relations, WL and WL+SL. Our
relations are significantly steeper, and have higher
normalizations\footnote[2]{Due to the addition of a third model parameter, the 
  CLASH normalization is not directly comparable to ours. However, visual
  inspection of Figure 3 shows that their value is certainly lower than ours.},
though it should be noted that we do not account for the lensing selection
function, which would lower both parameters.

\begin{figure*}
\begin{center}
\includegraphics[width=0.5\textwidth]{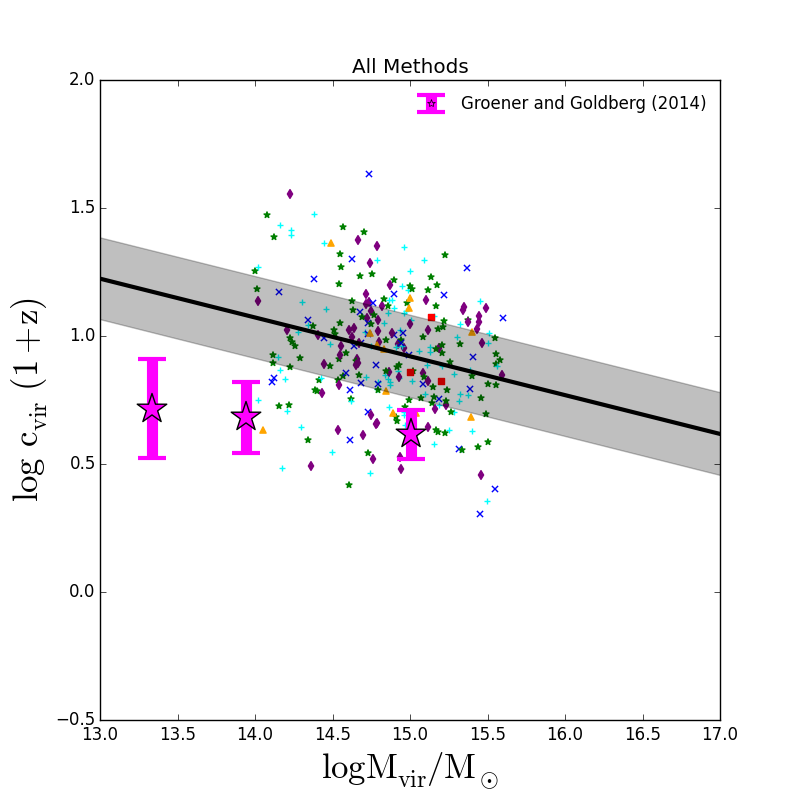}
\end{center}
\caption{The concentration-mass relation observed for the full cluster
  dataset. Color scheme is the same, though cyan data points represent co-added
  cluster measurements where more than one category of reconstruction method
  was used. Errorbars have been omitted here for clarity.}
\end{figure*}

\begin{figure*}
\begin{center}
\includegraphics[width=0.5\textwidth]{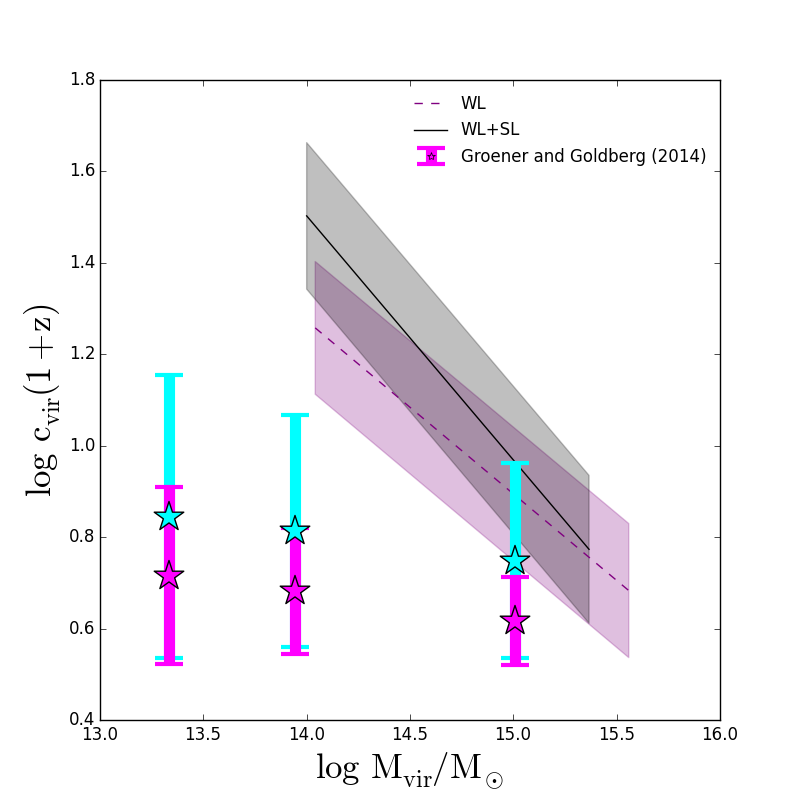}
\end{center}
\caption{WL and WL+SL relations plotted with MultiDark MDR1 Simulation results
  found by \citet{GR14.1}. Pink data points represent intrinsic 3D
  concentrations found in three mass bins, and cyan data points are
  corresponding 2D concentrations due to the projection of line-of-sight
  oriented halos.}
\end{figure*}

\begin{figure*}
\begin{center}$
\begin{array}{c}
\includegraphics[width=0.75\textwidth]{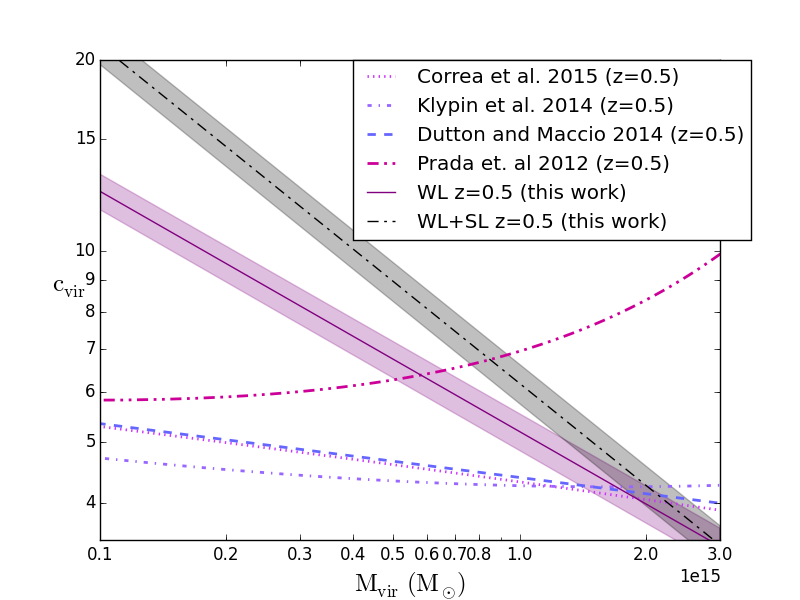}\\
  \includegraphics[width=0.75\textwidth]{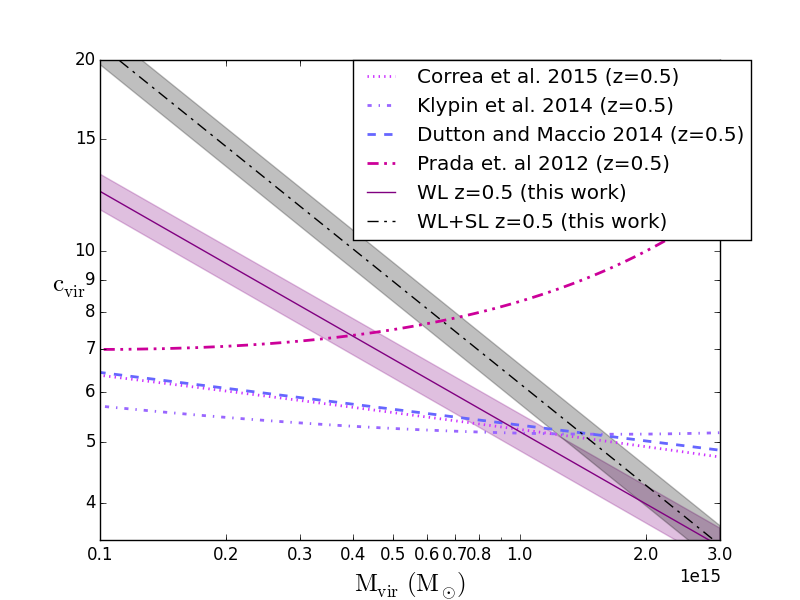}
\end{array}$
\end{center}
\caption{{\em Top:} Concentration-mass relations from recent simulations
  (\citealt{PR12.1}, \citealt{DU14.1}, \citealt{KL14.1}, and \citealt{CO15.2}), along with
  lensing (WL and WL+SL) relations found in this study. All relations are evaluated
at a redshift of $\mathrm{z=0.5}$. {\em Bottom:} Simulation relations after
projection effects have been taken into account. Halos are assumed to be
prolate spheroidal (q=0.65), oriented along the line-of-sight direction.}
\end{figure*}

\section{Projection, Shape, And A Direct Comparison Of Reconstruction Techniques}

When regarded as a single population of measurements, a linear fit to
the full dataset of cluster mass and concentration pairs can be said
to be, at face value, consistent with the results from simulations
(albeit only marginally). In Figure 4, we show the best-fit linear model to the
full dataset, with results from \cite{GR14.1} plotted in pink. We also find
good agreement with \citet{CO07.1}, who find a best-fit model of
$\mathrm{c_{vir} = \frac{14.8 \pm 6.1}{(1+z)} (M_{vir}/M_{*})^{-0.14\pm  0.12}}$.

When the projection of triaxial halos is taken into account, simulations become
more consistent with the lensing observations. Figure 5 compares WL and WL+SL
relations to intrinsic 3D halo concentrations (pink), and to 2D concentrations due to
line-of-sight projection (cyan) of MultiDark MDR1 simulation halos found
previously in \citet{GR14.1}. While projected halos in this figure represent a
perfectly elongated cluster sample, it is unlikely that {\em all} clusters with
lensing analyses performed to date are oriented in this way. Thus, projected
concentrations presented here can be interpreted as an upper limit, and
constrains the ability of line-of-sight projection in easing the tension between
simulations and lensing observations. \citet{BA12.1} also confirm that mock weak
lensing reconstructions of Millennium Simulation halos produce concentrations
of upwards of a factor of 2 for line-of-sight orientation, congruent with our
analytical treatment. However, this fails to completely account for the factor of
$\mathrm{\sim 3}$ ($\mathrm{\sim 4}$) which we find for WL (WL+SL) clusters
of mass $\mathrm{\sim 10^{14} M_{\odot}}$. 

In Figure 6, we compare our lensing relations to ones obtained through
dissipative {\em N-}body simulations found in the literature. Median simulation
relations are shown (top panel) over the mass range defined by our lensing samples
($\mathrm{1\times 10^{14} - 3\times 10^{15} M_{\odot}}$), and are
evaluated at a redshift corresponding to the average lensing redshift
($\mathrm{z=0.5}$) of our observational sample. Generally, the intrinsic scatter in
concentration is not shown here, but is assumed to follow a log-normal
distribution with a magnitude of $\mathrm{\Delta (\log c_{vir}) \sim 0.18}$
\citep{BU01.1}. The relation found by \citet{PR12.1} shows the prominent upturn
feature in concentration, while other relations are monotonically decreasing
functions of mass. Simulation relations and ones obtained in this study stand
in stark contrast with one another for lower mass clusters ($\mathrm{\lesssim
  1\times 10^{14} M_{\odot}}$), however, projection must be first be accounted
for before any conclusions can be drawn. In the bottom panel, we compare
analytical projections of simulation relations (using the method outlined in
\citealt{GR14.1}) with WL and WL+SL relations. For the purposes of
understanding the magnitude of this effect, halo shapes are assumed to be
well-described by prolate spheroidal isodensities with axis ratios of
$\mathrm{q=0.65}$ \citep{JI02.1}, with major axes in the line-of-sight
direction. Increased scatter in projected relations are expected to be caused
by the actual distributions of shapes and orientations (which we do not 
account for here). Direct statistical comparisons of these relations is
non-trivial, due to the differences in relation models. However, the projection of
triaxial halos was thought to be a sufficient explanation for fully describing
the existence of differing observed and simulated cluster concentrations. It
is clear that it is unlikely to be the sole contributing factor.

We also observe that the concentration-mass relation for combined WL+SL is
steeper than WL alone (though both relations are consistent at the
1-$\mathrm{\sigma}$ level). Cluster halo isodensities which are more prolate in
the inner regions can produce larger projected concentrations for line-of-sight
halos, and thus any method which makes use of information on this scale
may stand to be biased high because of it. We find that the sign of this
difference is in the right direction for this effect, and we cannot rule out
shape as one of the underlying causes.

Though we do not possess a complete volume-limited sample of galaxy clusters 
for which all measurement methods have been performed, we can begin to
understand any systematic effects present in clusters with concentrations and
masses present for various combinations. In Figure 7, we show clusters whose
profiles have been estimated using the following pairwise combinations of
methods: i) WL and WL+SL, ii) X-ray and WL, and iii) CM and LOSVD. We do not
detect any discernable trend in the way concentrations or masses are
overestimated or underestimated in each comparison, however, we show the
magnitude of the potential discrepancy. WL and WL+SL mass measurements are
generally in very good agreement with one another (with a few notable
exceptions), however, differences in concentration do exist which are upwards
of a factor of $\mathrm{\sim 2}$ in magnitude. X-ray and WL comparisons show
discrepancies in mass (concentration) which can reach as high as a factor of
$\mathrm{\sim 9}$ times ($\mathrm{\sim 6}$ times) larger, with X-ray mass
estimates tending to be larger than WL. Galaxy-based reconstruction
techniques (LOSVD/CM) tend to agree less in both mass and concentration, with
uncertainties which are quite large. 

\section{Conclusions And Discussions}
In this paper, we have studied the observed concentration-mass relation using
all known cluster measurements to date. We also model individual relations for
the most commonly used reconstruction techniques. In the present section, we
discuss our results of this study.

\begin{itemize}
\item There is an inconsistency between lensing (WL and WL+SL) concentrations
  and theoretical expectations from simulations. Low to medium mass lensing
  measurements ($\mathrm{\sim 10^{14} M_{\odot}}$) are inconsistent with
  simulation results, even when projection is taken into account. It is very
  likely the case that some of this difference can be generated by the
  existence of a strong orientation bias in the lensing cluster population,
  however, the magnitude of this effect (quantified by previous studies) cannot
  completely explain the difference we observe here.
\item We find that the concentration-mass relation from strong lensing
  clusters remains virtually unconstrained, due to the small size of the
  sample, as well as the insensitivity of SL reconstructions to the outer
  region of clusters.
\item The slope of the WL+SL relation is found to be higher (though still consistent)
  with WL alone over the lower half of the mass range, and may point to the existence
  of a new physical feature of clusters. However, when we only look at clusters
  with {\em both} measurements, we find no evidence that concentrations
  generated by WL+SL methods are in excess of WL. Most likely, this tells us
  that the selection effects for WL+SL is most likely the cause of this
  difference. Moreover, the intrinsic scatter of the concentration parameter on all
  mass scales is observed to be larger than the proposed difference in
  projected concentration due to shape, making this effect difficult to measure.
\item Lensing (WL and WL+SL) concentrations are systematically higher than
  those made with X-ray methods. In the mass range of $\mathrm{\sim (1-3)
    \times 10^{14} M_{\odot}}$, the WL+SL relation is marginally inconsistent
  with X-ray measurements. Reasons for a flatter X-ray relation as compared
  lensing methods are numerous. The gas distribution is rounder than the dark
  matter mass distribution, causing projection effects to 
  be less severe for X-ray samples. X-ray masses are also biased low due to
  temperature and hydrostatic equilibrium biases. Consequently, for the same
  nominal value of mass ($\mathrm{M_{WL} = M_{X}}$), X-ray clusters are likely
  more massive than clusters measured using WL. Because lower concentrations
  correlate with larger masses, lower concentrations are attributed to cluster
  mass bins, causing the X-ray c-M relation to have a lower normalization as
  compared to WL. Lastly, at very high masses, selection effects are less
  effective, since these clusters are likely to pass observational thresholds,
  and thus are included in samples. Due to less severe selection bias at larger
  mass ($\mathrm{\sim 10^{15} \,M_{\odot}}$), as well as lower concentrations
  as compared to WL, a bias towards flatness is expected for the X-ray c-M relation.
\item Out of all reconstruction methods, we also find that lensing (WL and
  WL+SL) relations are the {\em most} inconsistent with a power-law index of zero.
\item Methods which depend upon using galaxies as tracers of the mass
  show a neutral (LOSVD) or positive (CM) correlation between concentration and
  mass. The sensitivity of the slope of the caustic method c-M relation to the
  uncertainties is minimal. Disregarding uncertainties in either mass or
  concentration, we find a best fit slope and intercept of $\mathrm{m=0.207}$
  and $\mathrm{b=-2.103}$.
\item We find the c-M relation of our X-ray sample to be consistent with
  results from DM only simulations, though with a higher
  normalization, and slightly higher slope. However, direct comparison of these
  results with simulations  which include baryons, feedback, and star formation is
  necessary. \citet{RA13.1} performed such a study, and found that the
  dependence of the c-M relation on the radial range used to derive the
  relation, the baryonic physics included in simulations, and the selection of
  clusters based on X-ray luminosity all work to alleviate tensions between
  simulations and observations which existed previously. Though, they also find that
  including AGN feedback brings the relation more in line with DM only
  simulations, and it remains unclear whether or not all tensions between these
  relations have been identified and accounted for.
\end{itemize}

One potential source of error in the inference of the slope of the
c-M relation which we do not account for in this study is the covariance of the
mass and concentration measurements themselves \citep{SE15.2}. \citet{AU13.1}
discovered they were unable to constrain the slope of the c-M relation of a
sample of 26 strongly-lensed clusters with richness information, due to the
intrinsic covariance of their mass and concentration estimates, in addition to
a limited dynamic range of halo masses. Furthermore, improper modeling of the
distribution of halo masses can also significantly alter the inferred
relation (i.e. - it is sensitive to the prior). 

Selection effects can strongly steepen the slope of the c-M relation,
especially for lensing clusters \citep{ME14.1,ME14.2}. The slopes of relations
for clusters from CLASH, LOCUSS, SGAS, and a high redshift sample (also
included in this study), were all found to be much steeper than that of the
relation characterizing dark-matter only clusters \citep{SE15.2}. For fixed
mass, the most highly concentrated clusters are most likely to show SL
features, and thus are most likely to be included in SL selected samples
\citep{OG09.2}. In all cases, the selection process of clusters tend to prefer
over-concentrated halos, and depends strongly on observational selection
thresholds (Einstein radius, X-ray luminosity, morphology, etc.).

Another consideration is the mis-modeling of the halo profile. Recently, {\em
  N}-body simulations have shown that Einasto profiles provide an even more
accurate representation of the density profiles of dark matter halos compared
to the NFW profile \citep{DU14.1,KL14.1,ME14.2}. \citet{SE15.1} find that WL masses and
concentrations for very massive structures ($\mathrm{\gtrsim 10^{15} h^{-1}
  M_{\odot}}$) can be overestimated and underestimated, respectively, by about 
$\mathrm{\sim 10\%}$, if an NFW model is incorrectly assumed. Though this does
not fix the mismatch in the concentration parameter we have discussed here, it
could perhaps artificially steepen the overall slope of the relation by
reducing the concentrations of the most massive clusters.

Another plausible explanation for the existence of this new
over-concentration discrepancy for clusters is that dark matter only
simulations lack important cluster physics which is present in real
clusters. Feedback from AGN and supernovae, and gas cooling are mechanisms
which may cause (or prevent) further concentration of dark matter within the
cores of clusters, and have a strong effect on their lensing efficiency
\citep{PU05.1,WA08.1,RO08.1}. \citet{ME10.1} find that strong lensing
cross-sections for high mass clusters are boosted by up to 2-3 times, when
including gas cooling with star formation in simulations. Furthermore, they
find that by adding AGN feedback into the mix, this cross-section (and also the
concentration parameter) decreases, as energy is injected back into the
baryonic component.

There is a strong need to obtain low-mass ($\mathrm{< 1\times 10^{14}
  M_{\odot}}$) lensing measurements, since our most contentious conclusion is that, if the
  relation we have found holds in the galaxy group region, we expect cluster
  concentrations to be even less consistent with theory than they already
  are. Clearly this trend cannot continue indefinitely, but it remains to be
  seen how this model breaks down. An ideal study would contain a large,
  complete, and volume-limited sample of clusters, which can be studied in each
  reconstruction method. In this way, we could hope to eliminate the dependence
  of the selection function of clusters on the concentration-mass relation we
  would measure. Lastly, since selection effects are quite difficult to model, it is
  worth extending this study to as large of a sample as possible. Heterogeneous
  datasets (such as the one compiled in this study) have the ability to
  compensate for selection biases \citep{GO01.1,PI11.1,SE15.3}.

\bigskip

A.G. would like to thank the referee for a constructive and thorough review,
which significantly contributed to improving the quality of the publication.

M.S. acknowledges financial contributions from contracts ASI/INAF n.I/023/12/0
`Attivit\`a relative alla fase B2/C per la missione Euclid', PRIN MIUR
2010-2011 `The dark Universe and the cosmic evolution of baryons: from current
surveys to Euclid', and PRIN INAF 2012 `The Universe in the box: multiscale
simulations of cosmic structure'.

\bibliographystyle{mn2e_mod2}
\bibliography{paper2}

\newpage 

\begin{figure*}
\begin{center}$
\begin{array}{cc}
\includegraphics[width=0.7\textwidth]{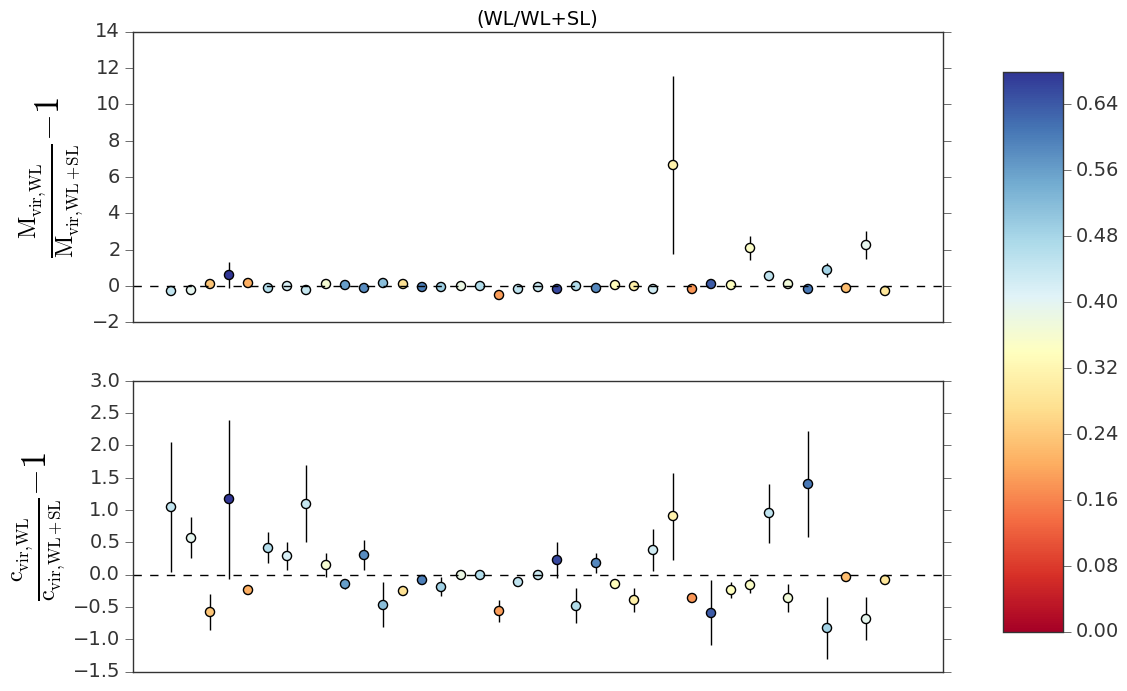} \\
  \includegraphics[width=0.7\textwidth]{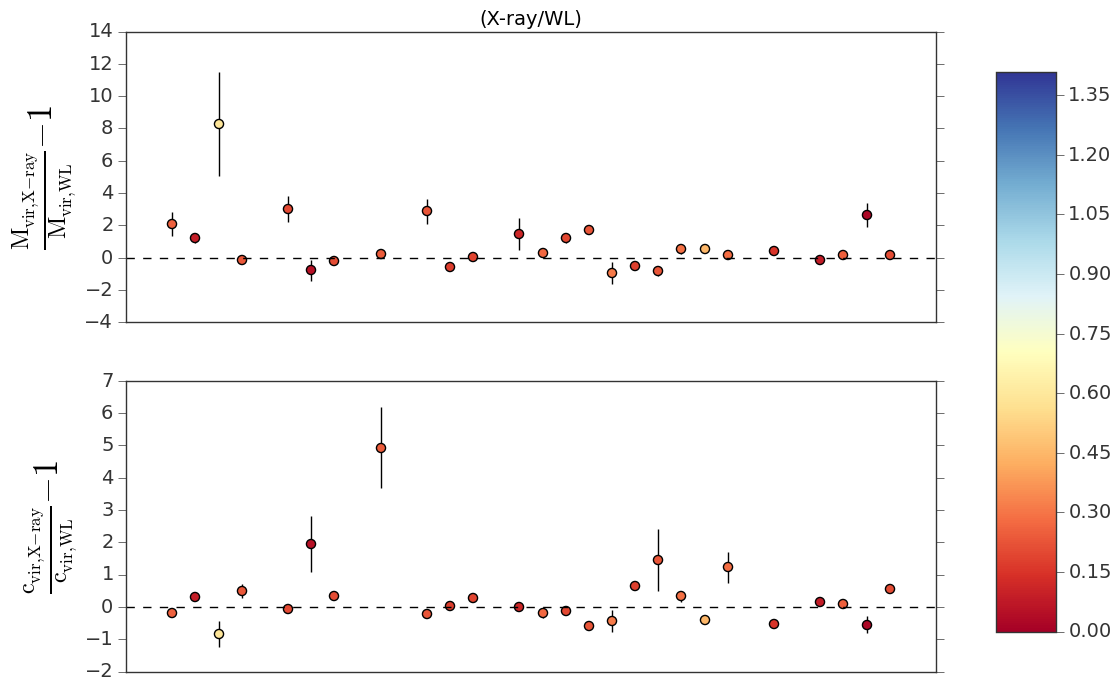} \\
\includegraphics[width=0.7\textwidth]{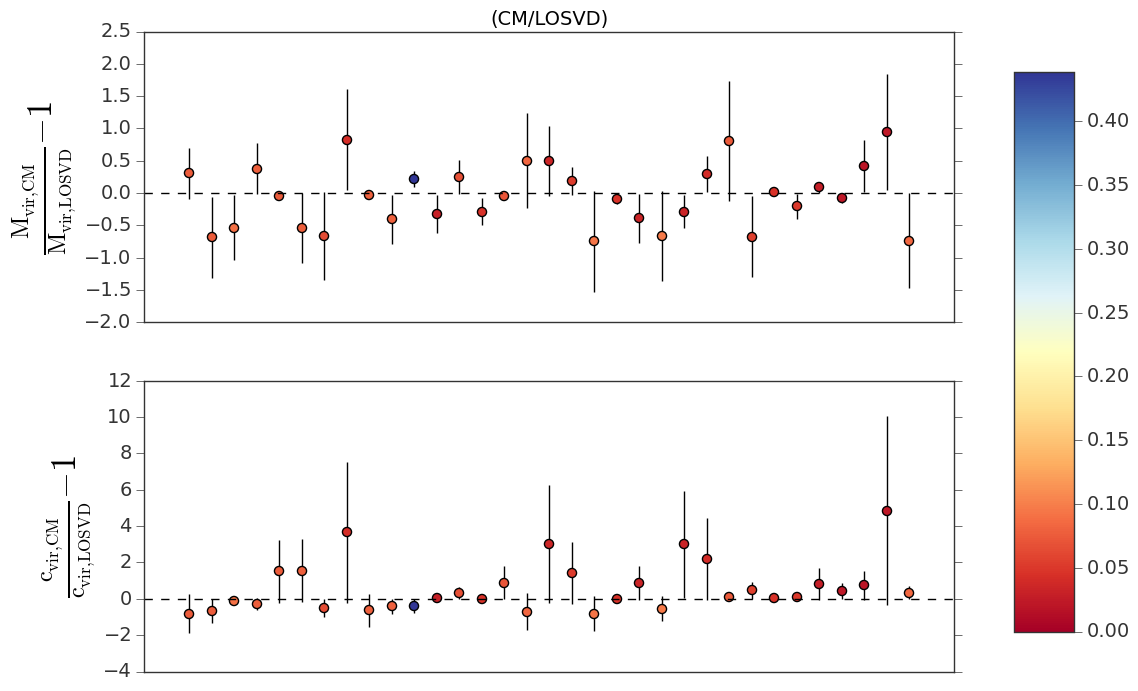}
\end{array}$
\end{center}
\caption{Comparisons of concentrations and masses for clusters measured
  in the following pairs of measurements: i) WL and WL+SL, ii) X-ray and WL,
  and iii) CM and LOSVD. In all cases, the color of the scatter point indicates
redshift.}
\end{figure*}



\begin{appendix}

\renewcommand{\thetable}{A-\arabic{table}}
\setcounter{table}{0}

\section*{Appendix A: Full Observational Dataset\footnotemark[1]}
We discuss here the details of our measurement aggregation procedure. 
\begin{itemize}
\item The overwhelming majority of measurements were reported in the one or
  both of the conventions shown in Table A-1 (200, and virial). Whenever
  possible, we report measurements made by the original paper, rather than
  relying on the conversion procedure outlined in \citet{HK03.1}. For papers
  which report their results for only one (or neither) of the previously
  mentioned conventions, we apply the aforementioned conversion process.
\item There are numerous definitions (and approximations) used throughout the
  literature for $\delta_{\mathrm{vir}}$ (also represented as
  $\Delta_{\mathrm{v}}$). All measurements reported using the virial
  overdensity convention have been converted to a consistent definition
  \citep{BR98.1}, before being reported in Table A-1:
\begin{equation}
\Delta_{\mathrm{v}} = 18\pi^{2} + 82x - 39x^{2}
\end{equation}
for a flat cosmology ($\Omega_{R} = 0$), and where $x = \Omega (z) -
1$. Furthermore, 
\begin{equation}
\Omega (z) = \frac{\Omega_{0} \left( 1+z \right)^{3}}{E(z)^{2}}
\end{equation}
with $E(z)$ representing the Hubble function.
\begin{equation}
E(z)^{2} = \Omega_{0} \left( 1+z \right)^{3} + \Omega_{R} \left( 1+z \right)^{2} + \Omega_{\Lambda}
\end{equation}
This approximation is accurate to 1\% within the range of $\Omega (z) =
0.1-1$.
\item All data reported in \citet{CO07.1}, were also reported in this study
  using their original cosmological model (with the exception of
  \citet{KI02.1}). We follow this convention, and continued to report
  measurements in Table A-1 in the cosmology found in the source paper.
\item All new measurements added to the dataset which do not appear in \citet{CO07.1}
  received redshifts from previous entries (if available; meaning that if the
  cluster already exists in the database, the first reported value of the
  redshift is used). Differences in these redshifts are minimal ($\mathrm{\sim
    1\%}$), and do not contribute significant uncertainty to the inferred c-M
  relation. Right ascension (RA) and declination (Dec) measurements were almost
  exclusively obtained from NED\footnotemark[2]. Lastly, due to the plurality of cluster
  naming conventions (nearly one for each survey or study), cluster names were
  cross-matched with previous entries using NED in order to ensure that our
  cluster sample does not contain artificially over-represented objects.
\end{itemize}

\footnotetext[1]{The raw data has been made publicly available (format: csv, xlsx)
  here: http://www.physics.drexel.edu/$\sim$groenera/}
\footnotetext[2]{https://ned.ipac.caltech.edu/}

\section*{Appendix B: Lensing Cosmology Correction}
In this section, we derive the correction to the measured cluster concentration and
mass (assuming an NFW profile), due to assumed cosmological model. Beginning
with the total mass enclosed within a sphere of radius $\mathrm{r}$
\begin{equation}
\mathrm{M_{NFW}(\le r) = 4\pi r_{s}^{3} \rho_{s} \left[ \log{(1+r/r_{s})} - \frac{r/r_{s}}{1+r/r_{s}} \right]}
\end{equation}
where $\mathrm{r_{s}}$ is the scale radius, and is used to scale the radial
coordinate which we will denote as $\mathrm{x = r/r_{s}}$. In terms of
projected quantities, following \citet{SE10.1} we can express the scale radius
and scale density $\mathrm{\rho_{s}}$ as
\begin{equation}
\mathrm{\rho_{s} = \frac{\Sigma_{cr}}{r_{s}} \kappa_{s}}
\end{equation}
\begin{equation}
\mathrm{r_{s} = D_{d} \theta_{s}}
\end{equation}
where $\mathrm{\kappa_{s}}$ is the normalization, $\mathrm{\Sigma_{cr}}$ is the
critical surface mass density for lensing, $\mathrm{D_{d}}$ is the angular
diameter distance to the lens, and $\mathrm{\theta_{s}}$ is the angular scale
radius.
\begin{equation}
\mathrm{\Sigma_{cr} = \frac{c^{2}D_{s}}{4\pi G D_{ds}D_{d}}}
\end{equation}
At this point it should be noted that the scale convergence and projected
(angular) scale radius do not depend upon cosmology when fitting the shear
profile. The mass within radius $\mathrm{r_{\Delta}}$ and its corresponding
concentration $\mathrm{c_{\Delta}}$ can be expressed in terms of projected
quantities 
\begin{equation}
\mathrm{M_{NFW}(\le r_{\Delta}) = \frac{c^{2}D_{d}D_{s} \kappa_{s}
  \theta_{s}^{2}}{G D_{ds}} \left[ \log{(1+c_{\Delta})} - \frac{c_{\Delta}}{1+c_{\Delta}} \right]}
\end{equation}
\begin{equation}
\mathrm{c_{\Delta} = \frac{r_{\Delta}}{r_{s}} = \frac{1}{D_{d}\theta_{s}} \left[
  \frac{2 M_{NFW}(\le r_{\Delta})}{\Delta \cdot H^{2}} \right]^{1/3}}
\end{equation}
where $\mathrm{\Delta}$ is the factor by which the density inside $\mathrm{r_{\Delta}}$ is
$\mathrm{\Delta \cdot \rho_{cr}}$, and $\mathrm{H}$ is the Hubble
parameter. Next, by solving the former two expressions for
$\mathrm{\kappa_{s}}$ and $\mathrm{\theta_{s}}$ (which are conserved
measurements for any arbitrary choice of cosmology), we obtain a system of
equations which then relate the lensing mass and concentration in any two cosmologies,
$\mathrm{\Omega_{1}}$ and $\mathrm{\Omega_{2}}$. In order to simplify the
notation a bit, the mass and concentration corresponding to $\mathrm{r_{\Delta}}$ 
in cosmology $\mathrm{\Omega_{x}}$, will henceforth be expressed as
$\mathrm{M_{\Delta}(\Omega_{x})}$ and $\mathrm{c_{\Delta}(\Omega_{x})}$,
respectively.

\begin{equation}
\mathrm{\frac{c_{\Delta}(\Omega_{2})^{3}}{M_{\Delta}(\Omega_{2})} =
  \frac{c_{\Delta}(\Omega_{1})^{3}}{M_{\Delta}(\Omega_{1})} \cdot R }
\end{equation}
\begin{equation}
\mathrm{\frac{f(c_{\Delta}(\Omega_{2}))}{M_{\Delta}(\Omega_{2})} =
  \frac{f(c_{\Delta}(\Omega_{1}))}{M_{\Delta}(\Omega_{1})} \cdot T }
\end{equation}

The ratios $\mathrm{R}$ and $\mathrm{T}$, and the function $\mathrm{f}$, can be
expressed in terms of the following cosmology dependent quantities:

\begin{equation}
\mathrm{R = \frac{D_{d}(\Omega_{1})^{3} H(\Omega_{1})^{2} \Delta
    (\Omega_{1})}{D_{d}(\Omega_{2})^{3} H(\Omega_{2})^{2} \Delta (\Omega_{2})}
  = \frac{D_{d}(\Omega_{1})^{3}}{D_{d}(\Omega_{2})^{3}} \cdot \frac{\Delta
    (\Omega_{1}) \rho_{cr}(\Omega_{1})}{ \Delta
    (\Omega_{2}) \rho_{cr}(\Omega_{1})} }
\end{equation}
\begin{equation}
\mathrm{T = \frac{D_{s}(\Omega_{1}) D_{d}(\Omega_{1})}{D_{ds}(\Omega_{1})}
  \frac{D_{ds}(\Omega_{2})}{D_{s}(\Omega_{2}) D_{d}(\Omega_{2})} =
  \frac{\Sigma_{cr}(\Omega_{2})}{\Sigma_{cr}(\Omega_{1})}}
\end{equation}
\begin{equation}
f(x) = \log(x) - \frac{x}{1+x}
\end{equation}

Solving this system of equations, can be done by numerically solving for
$\mathrm{c_{\Delta}(\Omega_{2})}$
\begin{equation} 
\mathrm{\frac{f(c_{\Delta}(\Omega_{2}))}{c_{\Delta}(\Omega_{2})^{3}} =
  \frac{f(c_{\Delta}(\Omega_{1}))}{c_{\Delta}(\Omega_{1})^{3}} \cdot \frac{T}{R} }
\end{equation}
Lastly, the mass $\mathrm{M_{\Delta}(\Omega_{2})}$ is obtained by direct
substitution of the numerical result from (15). 


\onecolumn
\begin{landscape}
\begin{center}
{\tiny 
\begin{longtable}{llllllllllll}
\caption{{\normalsize Cluster concentrations and masses}} \\
\hline \hline \\[-2ex]
   \multicolumn{1}{l}{Cluster} &
   \multicolumn{1}{l}{$\mathrm{z}$} &
   \multicolumn{1}{l}{$\mathrm{RA}$} &
   \multicolumn{1}{l}{$\mathrm{Dec.}$} &
   \multicolumn{1}{l}{Method} &
   \multicolumn{1}{l}{$\mathrm{c}_{200}$} &
   \multicolumn{1}{l}{$\mathrm{M}_{200}$} &
   \multicolumn{1}{l}{$\mathrm{c}_\mathrm{vir}$} &
   \multicolumn{1}{l}{$\mathrm{M}_\mathrm{vir}$} &
   \multicolumn{1}{l}{Ref.} &
   \multicolumn{1}{l}{$\delta$} &
   \multicolumn{1}{l}{$\Omega_{\mathrm{m}} / \Omega_{\Lambda} / \mathrm{h}$} \\
                               & & & & & &
\multicolumn{1}{l}{$(10^{14} \, \mathrm{M}_\odot)$} & &
\multicolumn{1}{l}{$(10^{14} \, \mathrm{M}_\odot)$} \\ [0.5ex] \hline
   \\[-1.8ex]
\endfirsthead
\multicolumn{8}{c}{{\tablename} \thetable{} -- {\it Continued}} \\[0.5ex]
  \hline \hline \\[-2ex]
   \multicolumn{1}{l}{Cluster} &
   \multicolumn{1}{l}{$\mathrm{z}$} &
   \multicolumn{1}{l}{$\mathrm{RA}$} &
   \multicolumn{1}{l}{$\mathrm{Dec.}$} &
   \multicolumn{1}{l}{Method} &
   \multicolumn{1}{l}{$\mathrm{c}_{200}$} &
   \multicolumn{1}{l}{$\mathrm{M}_{200}$} &
   \multicolumn{1}{l}{$\mathrm{c}_\mathrm{vir}$} &
   \multicolumn{1}{l}{$\mathrm{M}_\mathrm{vir}$} &
   \multicolumn{1}{l}{Ref.} &
   \multicolumn{1}{l}{$\delta$} &
   \multicolumn{1}{l}{$\Omega_{\mathrm{m}} / \Omega_{\Lambda} / \mathrm{h}$} \\
                               & & & & & &
\multicolumn{1}{l}{$(10^{14} \, \mathrm{M}_\odot)$} & &
\multicolumn{1}{l}{$(10^{14} \, \mathrm{M}_\odot)$} \\[0.5ex] \hline
  \\[-1.8ex]
\endhead
\hline \multicolumn{8}{l}{\textit{Continued on next page}} \\
\endfoot
\\[-1.8ex] \hline 
\endlastfoot
Virgo & 0.003 & 12 30 47.3 & +12 20 13 & X-ray & ${2.8}^{+0.7}_{-0.7}$ & ${4.2}^{+0.5}_{-0.5}$ & ${3.8}^{+0.9}_{-0.9}$ & ${5.4}^{+0.9}_{-0.9}$ & \citet{MC99.1} & 200 & 0.3/0.7/0.7 \\
Virgo & 0.003 & 12 30 47.3 & +12 20 13 & CM & ${0.92}^{}_{}$ & ${2.4}^{}_{}$ & ${1.35}^{}_{}$ & ${3.87}^{}_{}$ & \citet{RI06.1} & 200 & 0.3/0.7/None \\
NGC 4636 & 0.0031 & 12 42 49.8 & +02 41 16 & CM & ${8.64}^{}_{}$ & ${0.16}^{}_{}$ & ${11.41}^{}_{}$ & ${0.19}^{}_{}$ & \citet{RI06.1} & 200 & 0.3/0.7/None \\
NGC 5846 & 0.006 & 15 06 29.3 & +01 36 20 & CM & ${8.29}^{}_{}$ & ${0.5}^{}_{}$ & ${10.95}^{}_{}$ & ${0.59}^{}_{}$ & \citet{RI06.1} & 200 & 0.3/0.7/None \\
NGC 5044 & 0.009 & 13 15 24.0 & -16 23 08 & X-ray & ${8.4}^{+0.2}_{-0.2}$ & ${0.322}^{+0.009}_{-0.009}$ & ${11.1}^{+0.3}_{-0.3}$ & ${0.375}^{+0.011}_{-0.011}$ & \citet{GA06.1} & 1250 & 0.3/0.7/0.7 \\
Abell 1060 & 0.01 & 10 36 41.8 & -27 31 28 & LOSVD & ${10.67}^{+0.93}_{-2.52}$ & ${3.57}^{+0.34}_{-0.31}$ & ${13.98}^{+1.22}_{-3.31}$ & ${4.09}^{+0.39}_{-0.36}$ & \citet{WO10.1} & 102 & 0.3/0.7/0.7 \\
Abell 1060 & 0.01 & 10 36 41.8 & -27 31 28 & X-ray & ${10.27}^{+2.45}_{-2.17}$ & ${9.81}^{+1.32}_{-1.16}$ & ${13.7}^{+3.27}_{-2.89}$ & ${11.39}^{+1.53}_{-1.35}$ & \citet{BA14.1} & 200 & 0.27/0.73/0.73 \\
Abell 1060 & 0.01 & 10 36 41.8 & -27 31 28 & LOSVD & ${10.6}^{+17.1}_{-7.7}$ & ${3.8}^{+0.4}_{-0.7}$ & ${14.0}^{+22.0}_{-10.0}$ & ${4.4}^{+1.1}_{-1.0}$ & \citet{LO06.1} & virial & 0.3/0.7/0.7 \\
\end{longtable} }
\end{center}
\end{landscape}

\begin{center}
\begin{longtable}{|l|l|l|l|}
\caption{A Summary of The References}\\
\hline
Reference & {\footnotesize \# Measurements} & {\footnotesize \# Clusters (Unique)} & Method(s) \\
\hline
\endfirsthead
\multicolumn{4}{l}%
{\tablename\ \thetable\ -- \textit{Continued from previous page}} \\
\hline
Reference & {\footnotesize \# Measurements} & {\footnotesize \# Clusters (Unique)} & Method(s) \\
\hline
\endhead
\hline \multicolumn{4}{l}{\textit{Continued on next page}} \\
\endfoot
\hline
\endlastfoot
{\footnotesize \citet{BA14.1}} & 128 & 128 & {\small X-ray} \\
{\footnotesize \citet{SE14.1}} & 109 & 104 & {\small WL} \\
{\footnotesize \citet{RI06.1}} & 72 & 72 & {\small CM} \\
{\footnotesize \citet{OG12.1}} & 56 & 28 & {\small WL+SL},\,{\small WL} \\
{\footnotesize \citet{ET11.1}} & 44 & 44 & {\small X-ray} \\
{\footnotesize \citet{WO10.1}} & 41 & 41 & {\small LOSVD} \\
{\footnotesize \citet{SC07.1}} & 31 & 31 & {\small X-ray} \\
{\footnotesize \citet{OK10.1}} & 26 & 26 & {\small WL} \\
\end{longtable}
\end{center}

\end{appendix}

\end{document}